\documentclass{elsart}
\usepackage{epsfig}
\usepackage{enumerate}

\begin{document}
\begin{frontmatter}

\title{Consequence of reputation in the Sznajd consensus model}
\author{Nuno Crokidakis$^{1,2}$}
\thanks{nuno@if.uff.br}
\author{and Fabricio L. Forgerini$^{2,3}$}
\thanks{fabricio$\_$forgerini@ufam.edu.br}
\address{
$^{1}$Instituto de F\'{\i}sica - Universidade Federal Fluminense \\
Av. Litor\^anea s/n \\
24210-340 \hspace{5mm} Niter\'oi - RJ \hspace{5mm} Brazil \\
$^{2}$Departamento de F\'{\i}sica, I3N - Universidade de Aveiro \\
3810-193 \hspace{5mm} Aveiro \hspace{5mm} Portugal \\
$^{3}$ISB - Universidade Federal do Amazonas \\
69460-000 \hspace{5mm} Coari - AM \hspace{5mm} Brazil}

\maketitle

\begin{abstract}
\noindent
In this work we study a modified version of the Sznajd sociophysics model. In particular we introduce reputation, a mechanism that limits the capacity of persuasion of the agents. The reputation is introduced as a score which is time-dependent, and its introduction avoid dictatorship (all spins parallel) for a wide range of parameters. The relaxation time follows a log-normal-like distribution. In addition, we show that the usual phase transition also occurs, as in the standard model, and it depends on the initial concentration of individuals following an opinion, occurring at a initial density of up spins greater than $1/2$. The transition point is determined by means of a finite-size scaling analysis.

\end{abstract}
\end{frontmatter}


\section{Introduction}

In the last few years, the Sznajd sociophysics model \cite{sznajd_model} has been successfully applied to many different areas, like politics, marketing and finance (for reviews, see \cite{stauffer_review,sznajd_review} and more recently \cite{loreto_rmp}). The main goal of the model is the emergence of consensus from simple microscopic rules based on an Ising-type model. After the introduction of the original one-dimensional model in 2000 \cite{sznajd_model}, many modifications were proposed, like the extension to the square \cite{adriano}, triangular \cite{chang} and cubic lattices \cite{bernardes}, the increase of the range of the interaction \cite{schulze1} and the number of variable's states \cite{stauffer_adv,sznajds,bonnekoh}, adding diffusion \cite{stauffer_adv,schulze2}, and others.

The Sznajd model defined on the square lattice was firstly studied by Stauffer \textit{et al.} \cite{adriano}. Considering that a $2\times 2$ plaquette with all spins parallel can convince their eight neighbors (we can call this Stauffer's rule), the authors found a phase transition for initial density of up spins $d=1/2$. The two-dimensional model was also considered in a randomly diluted square lattice \cite{andre}, where every site may carries one spin or is empty. A situation with long-range correlations between the occupation of various lattice sizes was also considered, but in both cases the results are very similar to those on regular lattices \cite{stauffer_review}. A more realistic situation is to consider a probability of persuasion. The Sznajd model is robust against this situation: if one convinces the neighbors only with some probability $p$, and leaves them unchanged with probability $1-p$, still a consensus is reached after a long time \cite{stauffer_review}. Models that consider many different opinions (using Potts' spins, for example) or defined on small-world networks were studied in order to represent better approximations of real communities' behavior (see \cite{stauffer_review} and references therein). In order to avoid full consensus in the system, and makes the model more realistic, Schneider introduced opportunists and persons in opposition \cite{schneider}.

Unfortunately the dynamics of social relationships in the real world shows a large number of details that are commonly neglected in some theoretical models. In order to introduce a more realistic feature, we have considered in this work a reputation mechanism that limits the capacity of persuasion of the agents. It is expected that the inclusion of reputation in the Sznajd model turns it closer to a real social system, where not only the number of individuals with same opinion matters. We believe that the reputation of the agents who holds same opinion is an important factor in persuasion across the community. In other words, it is realistic to believe that individuals change their opinions if they are influenced by high-reputation persons. In fact, we show that a democracy-like situation, with a ferromagnetic ordering with not all spins parallel, is possible considering simple microscopic rules.

This Letter is organized as follows. In Section 2 we present the model and define their microscopic rules. The numerical results as well as the finite-size scaling analysis are discussed in Section 3. Our conclusions are presented in Section 4.


\section{Model}

We have considered the Sznajd model defined on a square lattice with linear size $L$ (i.e., $L^{2}$ \textit{agents}). Our model is based on the Stauffer's rule (rule Ia of the model in \cite{adriano}). Moreover, an integer number ($R$) labels each player and represents its reputation across the community, in analogy to the Naming game model considered by Brigatti \cite{edgardo}. The reputation is introduced as a score which is time-dependent. The agents start with a random distribution of the $R$ values, and during the time evolution, the reputation of each agent changes according to its capacity to be persuasive, following the rules explained below.

At each time step, the following microscopic rules control our model:
\begin{enumerate}
\item We randomly choose a 2 $\times$ 2 plaquette of four neighbors.
\item If not all four center spins are parallel, leaves its eight neighbors unchanged.
\item If all four center spins are parallel, we calculate the average reputation of the plaquette:
\begin{eqnarray} \nonumber
{\rm average\; reputation}= \frac{1}{4}\sum_{i=1}^{4}R_{i}~,
\end{eqnarray}
where each term $R_{i}$ represents the reputation of one of the plaquettes' agents \footnote{Notice that we will consider the integer part of the ratio $\sum_{i=1}^{4}R_{i}/4$.}.
\item In this case, we compare the reputations of each one of the eight neighbors with the average reputation. If the reputation of a neighbor is less than the average one, this neighbor follows the plaquette orientation. For each persuaded individual, each one of the plaquette agents' increase their reputations by 1 (in other words, the average reputation of the plaquette is increased by 1).
\end{enumerate}

Thus, even in the case where all plaquettes' spins are parallel, a different number of agents may be convinced, namely, 8, 7, 6, ..., 1 or 0 ones. As pointed by Stauffer \cite{stauffer_review}, we can imagine that each agent in the Sznajd model carries an opinion, that can either be up (e.g. Republican) or down (e.g. Democrat) and that represents one of two possible opinions on any question. The objective of the agents in this \textit{game} is to convince their neighbors of their opinion. One can expect that, if a certain group of agents convince many other agents, their persuasion abilities increase. Thus, the inclusion of reputation in the model may capture this real-world characteristic.


\section{Numerical Results}

\begin{figure}[t]
\begin{center}
\includegraphics[width=0.45\textwidth,angle=0]{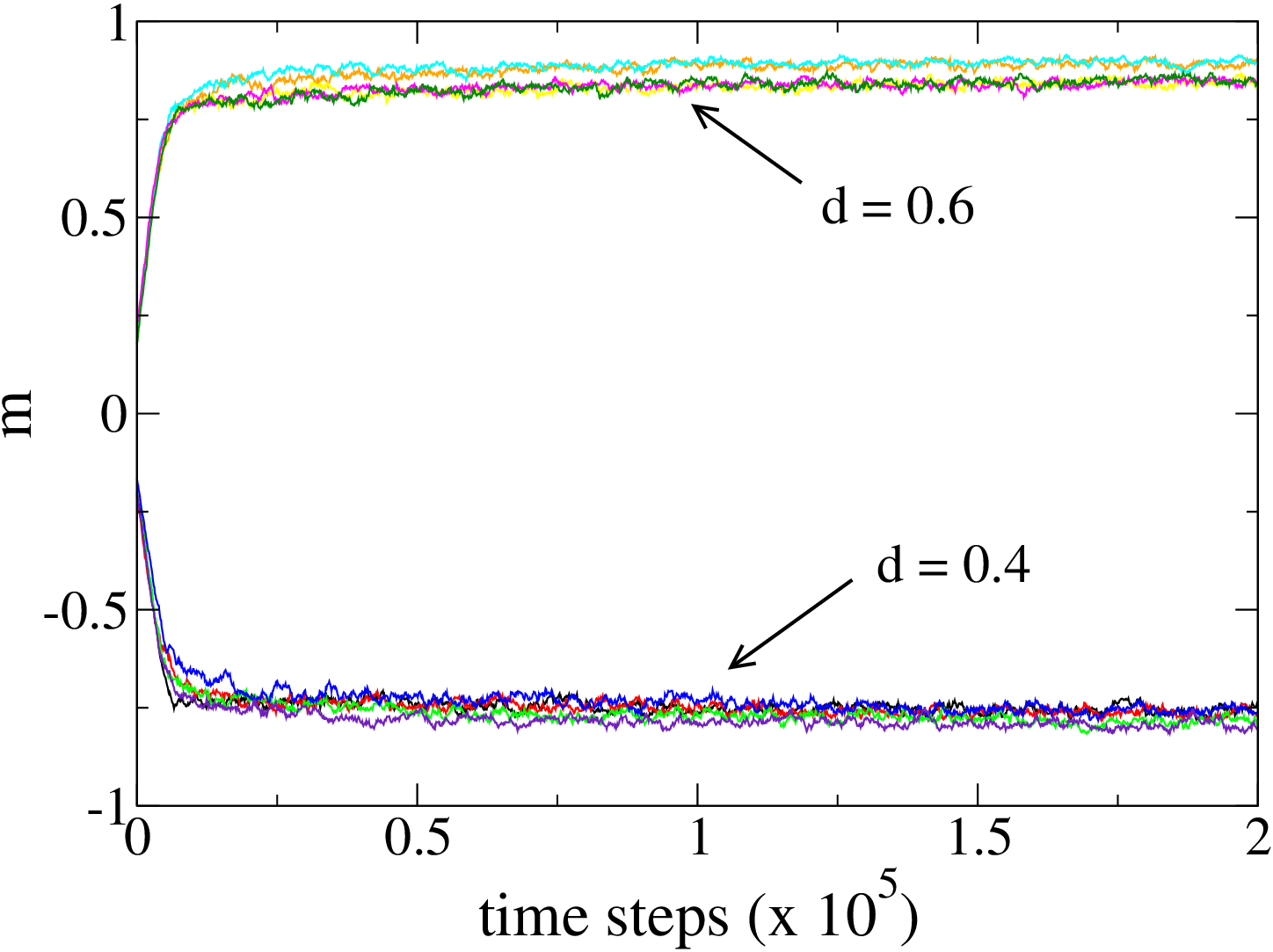}
\hspace{0.4cm}
\includegraphics[width=0.44\textwidth,angle=0]{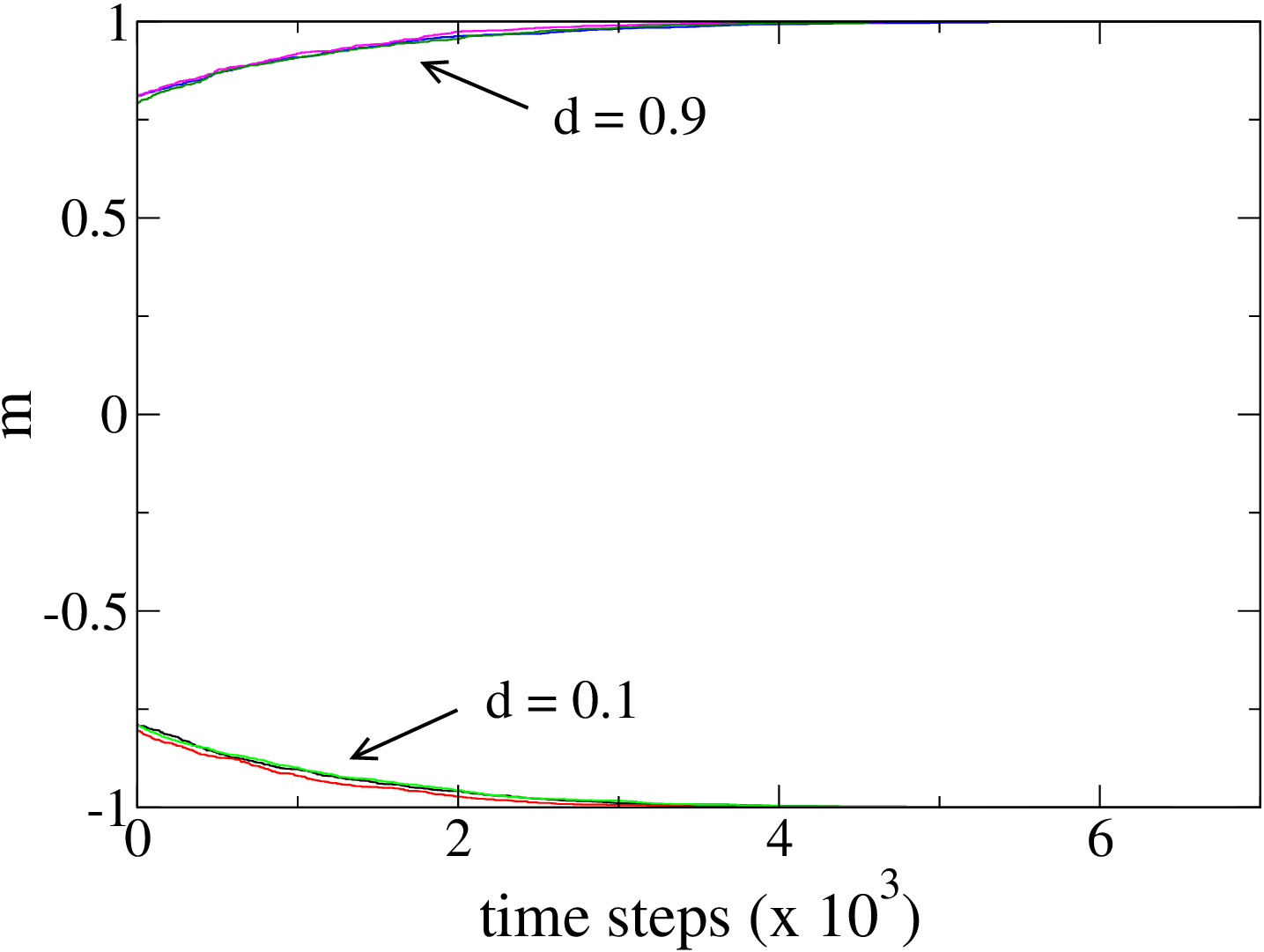}
\end{center}
\caption{(Colour online) Time evolution of the magnetization for $L=53$, initial densities of up spins $d=0.4$ and $d=0.6$ and different samples (left side). We can see that the steady states show situations where the total consensus is not obtained, in opposition of the standard Sznajd model defined on the square lattice \cite{adriano}. In the right side we show the results for $d=0.1$ and $d=0.9$. In these cases the system reaches consensus.}
\label{Fig1}
\end{figure}

In the simulations, the initial values of the agents' reputation follow a gaussian distribution centered at $0$ with standard deviation $\sigma=5$ \footnote{In the following, we discuss the effects of increasing the standard deviation of the initial distribution of agents' reputation.}. Following the previous works on the Sznajd model, we can start studying the time evolution of the magnetization,
\begin{equation}
m=\frac{1}{N}\sum_{i=1}^{N}s_{i}~,
\end{equation}
\noindent
where $N=L^{2}$ is the total number of agents and $s_{i}=\pm 1$. In the standard Sznajd model defined on the square lattice \cite{adriano}, the application of the Stauffer's rule, where a $2\times 2$ plaquette convince its eight neighbors if all spins are parallel, with initial density of up spins $d=1/2$ leads the system to the fixed points with all up or all down spins with equal probability. For $d<1/2$ ($>1/2$) the system goes to a ferromagnetic state with all spins down (up) in all samples, which characterizes a phase transition at $d=1/2$ in the limit of large $L$. As pointed by the authors in \cite{adriano}, fixed points with all spins parallel describe the published opinion in a dictatorship, which is not a commom situation nowadays. However, ferromagnetism with not all spins parallel corresponds to a democracy, which is very commom in our world. We show in Fig. \ref{Fig1} the behavior of the magnetization as a function of time in our model, considering the above-mentioned rules. In the left side of Fig. \ref{Fig1}, we show a value of $d>1/2$ ($<1/2$), and one can see that the total consensus with all spins up (down) will not be achieved in any sample, indicating that (i) a democracy-like situation is possible in the model without the consideration of a mixing of different rules \cite{adriano}, or some kind of special agents, like contrarians and opportunists \cite{schneider}, and (ii) if a phase transition also occurs in our case, the transition point will be located somewhere at $d>1/2$. Also in Fig. \ref{Fig1} (right side), we show situations where the consensus is obtained with all spins up (for $d=0.9$) and with all spins down (for $d=0.1$).

\begin{figure}[t]
\begin{center}
\includegraphics[width=0.45\textwidth,angle=0]{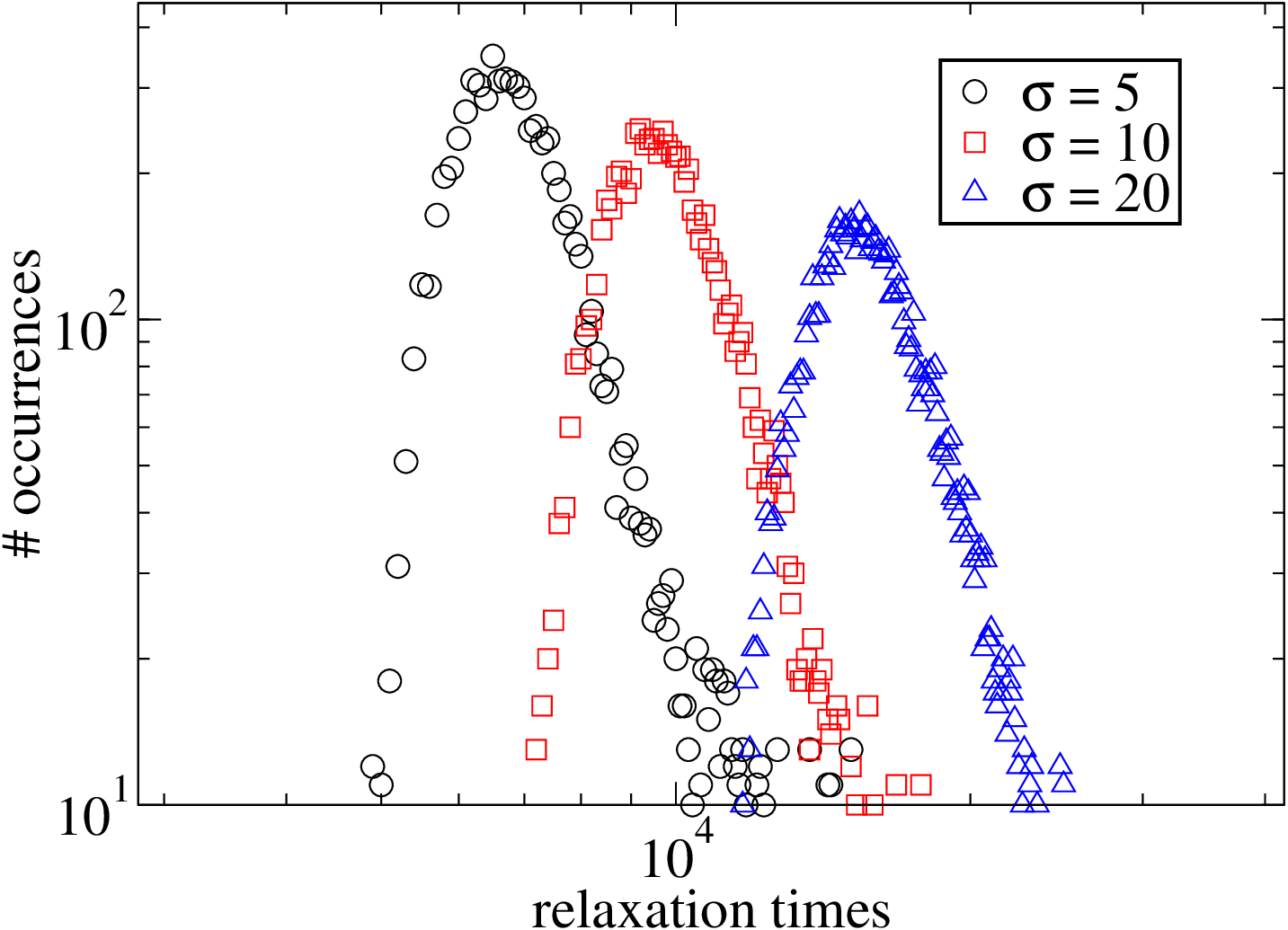}
\hspace{0.4cm}
\includegraphics[width=0.45\textwidth,angle=0]{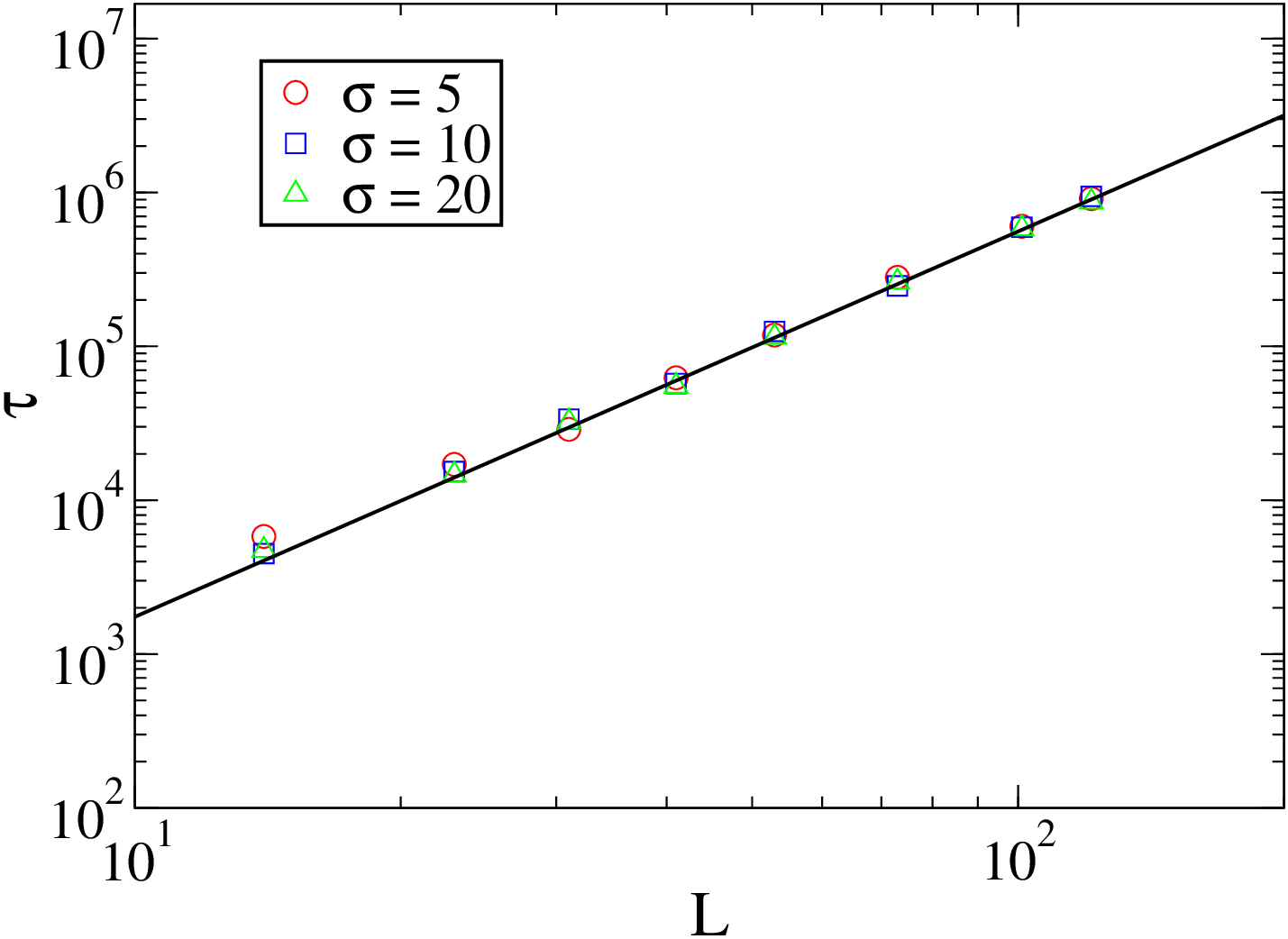}
\end{center}
\caption{(Colour online) Left side: log-log plot of the histogram of relaxation times for $L=53$ and $d=0.8$, obtained from $10^{4}$ samples, with agents' initial reputations following a gaussian distribution with different standard deviations $\sigma$. The distribution is compatible with a log-normal one for all values of $\sigma$, which corresponds to the observed parabola in the log-log plot. Right side: average relaxation time $\tau$, over $10^{4}$ samples, versus latice size $L$ in the log-log scale. The straight line has slope 5/2. The result is robust with respect to the choice of different $\sigma$ values.}
\label{Fig2}
\end{figure}

\begin{figure}
\begin{center}
\vspace{0.5cm}
\includegraphics[width=0.44\textwidth,angle=0]{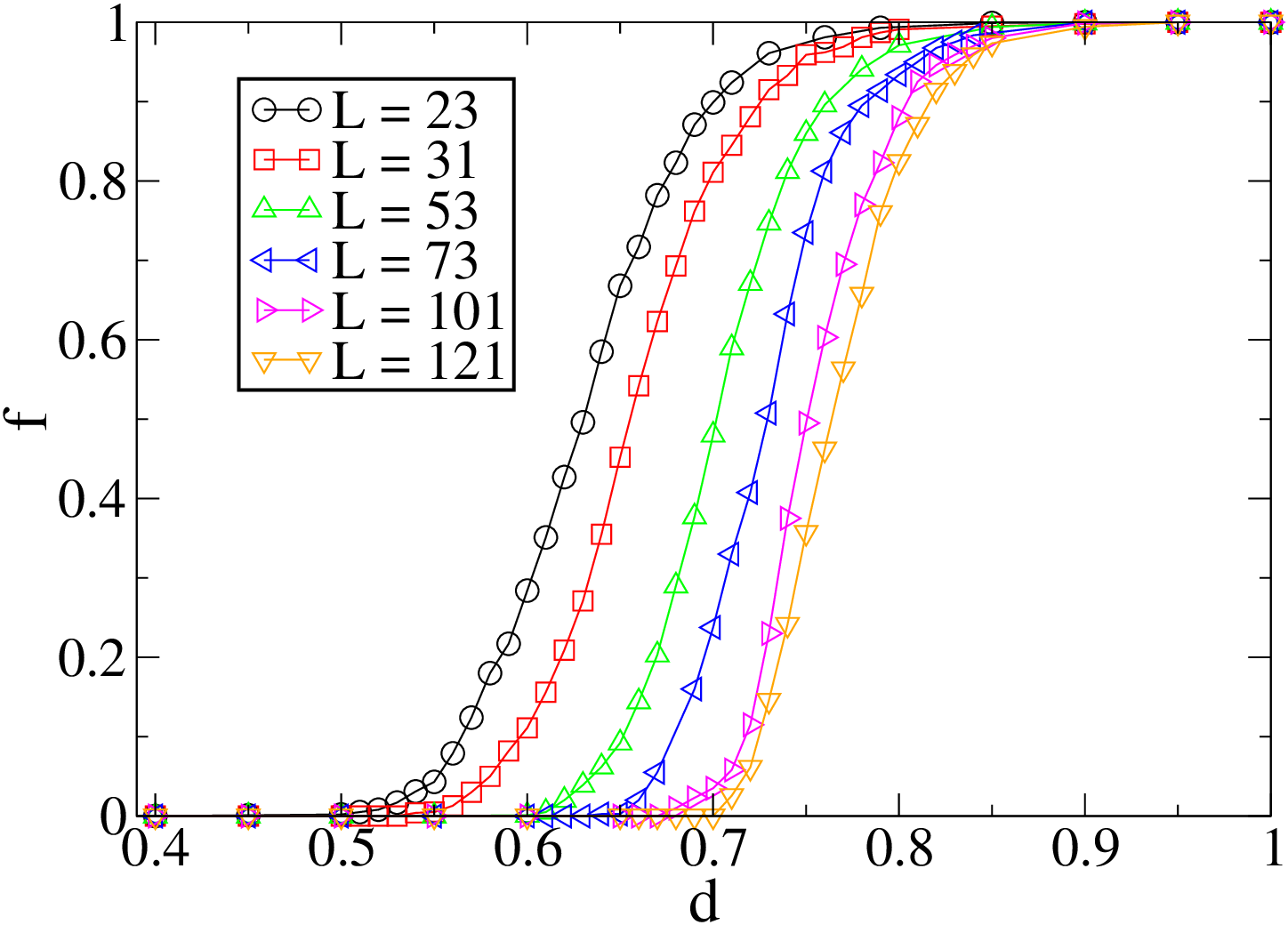}
\hspace{0.4cm}
\includegraphics[width=0.44\textwidth,angle=0]{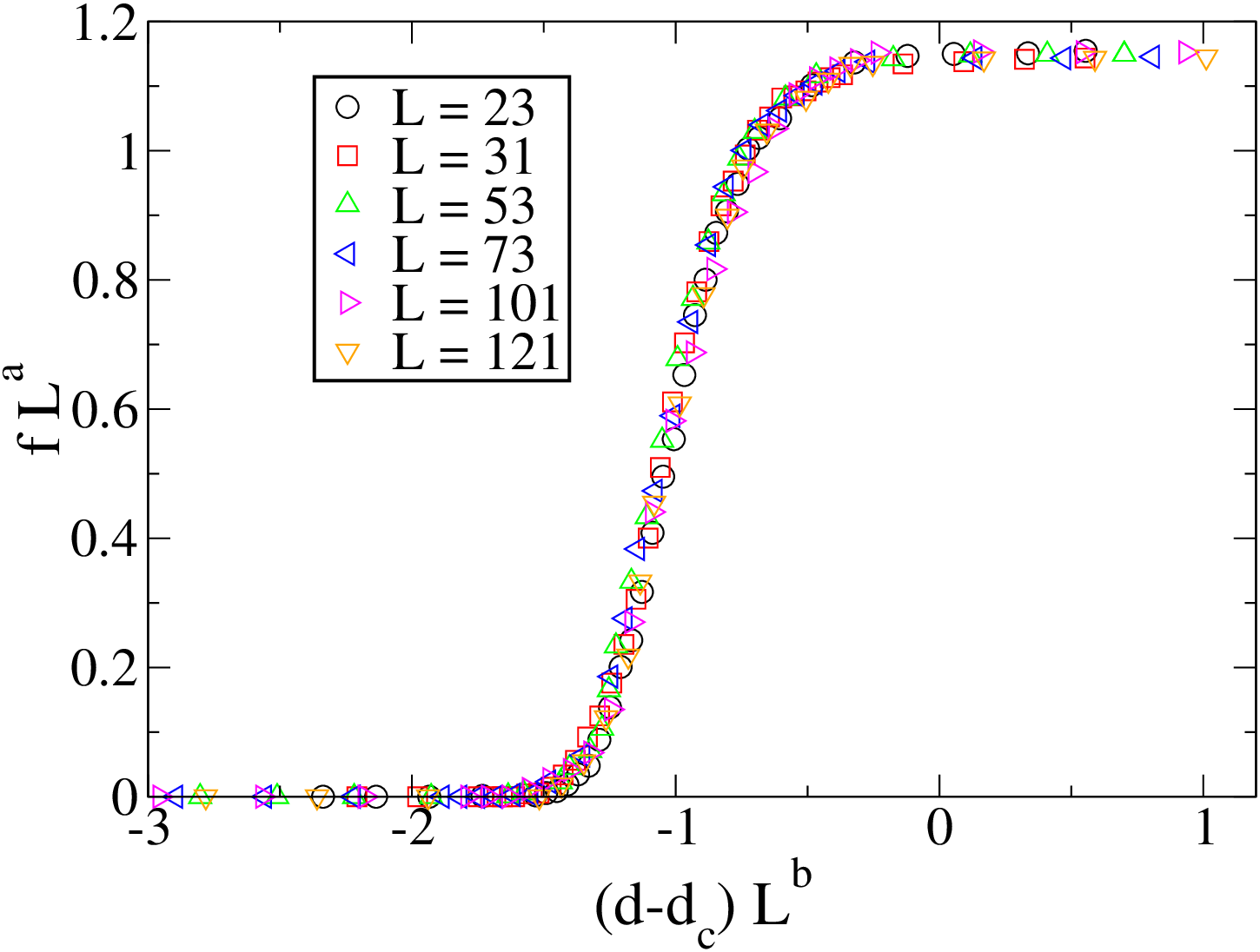}
\end{center}
\caption{(Colour online) Left side: fraction $f$ of samples which show all spins up when the initial density $d$ is varied in the range $0.4<d<1.0$, for some lattice sizes $L$. The total number of samples is $1000$ (for $L=23$, $31$ and $53$), $500$ (for $L=73$) and $200$ (for $L=101$ and $121$). Right side: the corresponding scaling plot of $f$. The best collapse of data was obtained for $a=0.035\pm 0.002$, $b=0.444\pm 0.002$ and $d_{c}=0.88\pm 0.01$.}
\label{Fig3}
\end{figure}

As in the previous studies of the Sznajd model \cite{sznajd_model,stauffer_review,adriano,adriano2}, the relaxation time, i.e., the time needed to find all the agents at the end having the same opinion, depends on the model's parameters. The distribution of the number of sweeps through the lattice, averaged over $10^{4}$ samples, needed to reach the fixed point is shown in Fig. \ref{Fig2} (left side). We can see that the relaxation time distribution is compatible with a log-normal one for all values of the standard deviation $\sigma$, which corresponds to a parabola in the log-log plot of Fig. \ref{Fig2} (left side). The same behavior was observed in other studies of the Sznajd model \cite{adriano,schneider,adriano2}. In the right side of Fig. \ref{Fig2} we show the average relaxation time $\tau$ (also over $10^{4}$ samples) versus latice size $L$ in the log-log scale. We can verify a power-law relation between these quantities in the form $\tau\sim L^{5/2}$, for all values of the standard deviation, which indicates that the result is robust with respect to the choice of different $\sigma$ values. Power-law relations between $\tau$ and $L$ were also found in a previous work on the model \cite{adriano2}.

\begin{figure}[t]
\begin{center}
\includegraphics[width=0.5\textwidth,angle=0]{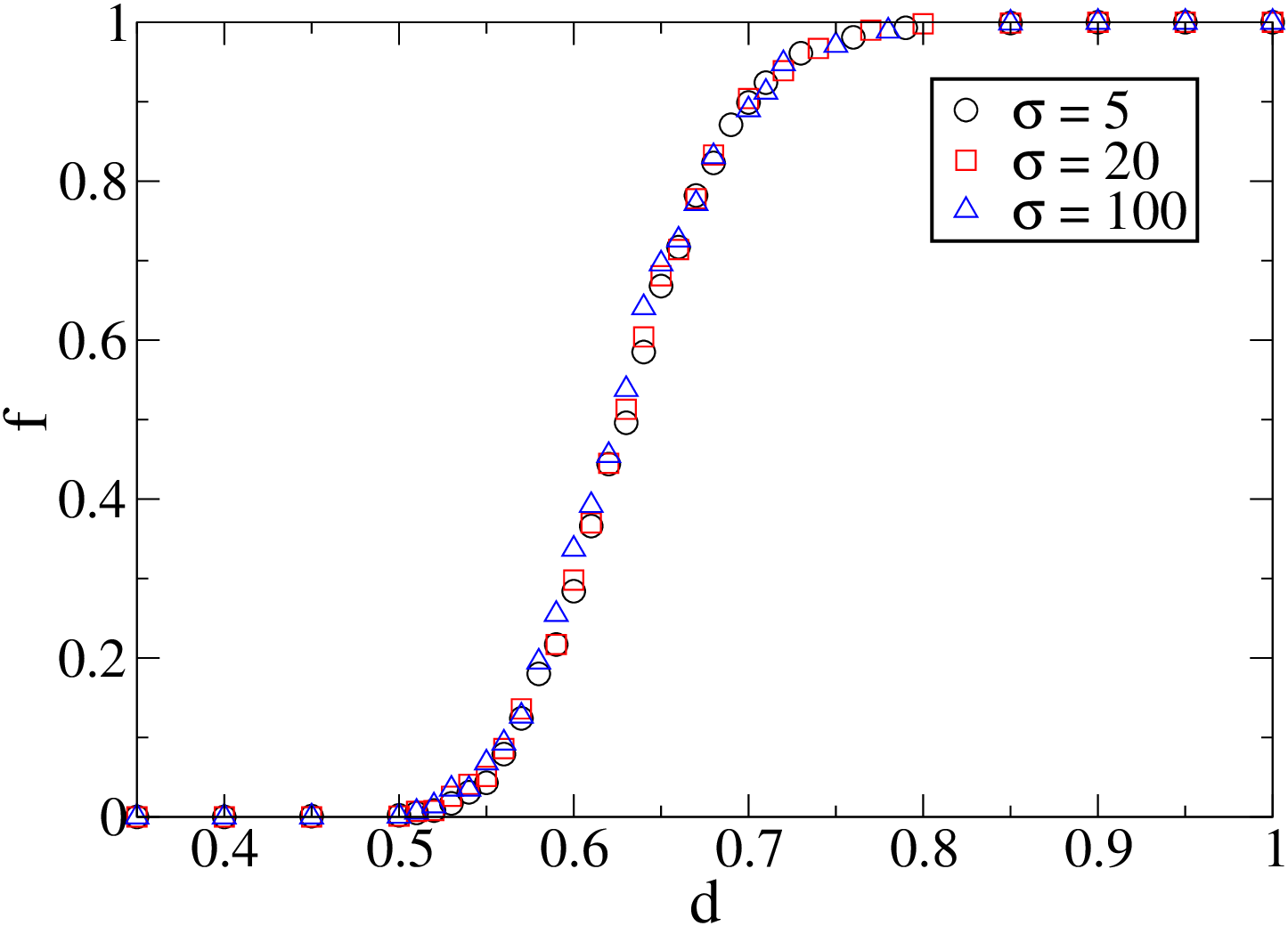}
\end{center}
\caption{(Colour online) Fraction $f$ of samples which show all spins up when the initial density $d$ is varied in the range $0.35<d<1.0$, for $L=23$, $1000$ samples and some different values of $\sigma$. This result show that the increase of $\sigma$ do not change the behavior of $f$.}
\label{Fig4}
\end{figure}

In order to analyze in more details the phase transition, we have simulated the system for different lattice sizes $L$ and we have measured the fraction of samples which show all spins up when the initial density of up spins $d$ is varied in the range $0.4<d<1.0$. We have considered $1000$ samples for $L=23$, $31$ and $53$, $500$ samples for $L=73$ and $200$ samples for $L=101$ and $121$. The results are shown in Fig. 3 (left side). One can see that the transition point is located somewhere in the region $d>1/2$, as above discussed. In order to locate the critical point, we performed a finite-size scaling (FSS) analysis, based on the standard FSS equations \cite{adriano2},
\begin{eqnarray}
f(d,L) & = & L^{-a}\;\tilde{f}((d-d_{c})\;L^{-b}) ~, \\
d_{c}(L) & = & d_{c}+a\;L^{-b} ~.
\end{eqnarray}
\noindent
 The result is shown in Fig. 3 (right side), and we have found that 
\begin{equation}
d_{c}=0.88 \pm 0.01~,
\end{equation}
\noindent
in the limit of large L. The critical point occurs at $d>1/2$, different of the Sznajd model without reputation defined on the square lattice. This fact may be easily understood: at each time step, the randomly chosen 2$\times$2 plaquette may convince 8, 7, 6, ..., 1 or 0 neighbors, even if the plaquettes' spins are parallel. In the standard model, if the plaquettes spins' orientations are the same, 8 neighbors are convinced immediately, thus it is necessary a smaller initial density of up spins to the system reaches the fixed point with all spins up. However, the existence of a critical value $d_{c}<1.0$ indicates that a phase transition also occurs in our version of the Sznajd model. This result is robust with respect to the choice of different values of $\sigma$ (see Fig. \ref{Fig4}).


\section{Conclusions}

In this work, we have analyzed the effects of the introduction of agents' reputation in the square lattice Sznajd model. The reputation is introduced as a score ($R$) which is time-dependent. The agents start with a gaussian distribution of the $R$ values, and during the time evolution, the reputation of each agent changes according to its capacity of persuasion, following the model's rules. We expect that the consideration of agents' reputation makes the Sznajd model more realistic. In fact, take into account simple microscopic rules, a democracy-like situation emerges spontaneously in the model, for a wide range of the initial densities $d$ of up spins. The consensus with all spins up (down) is obtained only for large (small) values of $d$. 

We performed Monte Carlo simulations on square lattices with linear sizes up to $L=121$ and typically $10^{3}-10^{4}$ samples. As in the standard model \cite{adriano}, we found a log-normal distribution of the relaxation times. In addition, the average relaxation time $\tau$ depends on the lattice size in a power-law form, $\tau\sim L^{5/2}$, which is independent of the standard deviation $\sigma$ of the gaussian distribution of the agents' initial reputation. The system also undergoes a phase transition, as in the traditional model, but the critical initial density of up spins was found to be $d_{c}=0.88$, in the limit of large lattices, greater than $1/2$, the value found by Stauffer \textit{et al.} \cite{adriano}. This fact may be easily understood: at each time step, the randomly chosen 2$\times$2 plaquette may convince 8, 7, 6, ..., 1 or 0 neighbors, even if the plaquettes' spins are parallel. In the standard case, if the plaquettes spins' orientations are the same, 8 neighbors are convinced immediately, thus it is necessary a smaller initial density of up spins to the system reaches the fixed point with all spins up. The simulations indicate that the observed phase transition is robust with respect to the choice of different values of $\sigma$.

As extensions of this work, it could be interesting to explore the role of the system topology. Different complex topologies could be studied for agents embedded on more realistic networks, like small-world ones. Agents with more than two opinions, using Potts' spins, for example, can also be interesting to study.

\section*{Acknowledgments}

The authors are grateful to Edgardo Brigatti and Adriano O. Sousa for discussions about social models. N. Crokidakis would like to thank the Brazilian funding agency CAPES for the financial support at Universidade de Aveiro at Portugal. Financial support from CNPq is also acknowledge. F. L. Forgerini would would like to thank the ISB - Universidade Federal do Amazonas for the support.

\end{document}